\documentclass[9pt,journal,compsoc]{IEEEtran}

\ifCLASSOPTIONcompsoc
  \usepackage[nocompress]{cite}
\else
  \usepackage{cite}
\fi
\ifCLASSINFOpdf
\else
\fi

\usepackage{fancyhdr}
\usepackage{hyperref}

\usepackage[dvipsnames]{xcolor}
\usepackage{comment}
\usepackage{xspace}
\usepackage[linesnumbered,ruled]{algorithm2e}
\usepackage{cite}
\usepackage{booktabs}     
\usepackage{amsfonts,amssymb,amsmath}  
\usepackage{nicefrac}     
\usepackage[flushleft]{threeparttable}
\usepackage{multirow}
\usepackage{tikz}
\usepackage{url}

\usepackage{textcomp}

\newcommand{\beas}{\begin{eqnarray*}}
\newcommand{\eeas}{\end{eqnarray*}}
\newcommand{\bea}{\begin{eqnarray}}
\newcommand{\eea}{\end{eqnarray}}
\newcommand{\bes}{\begin{equation*}}
\newcommand{\ees}{\end{equation*}}
\newcommand{\be}{\begin{equation}}
\newcommand{\ee}{\end{equation}}

\iftrue
\newcommand{\nskim}[1]{{\color{red}[\textbf{\sc nskim}: \textit{#1}]}}
\newcommand{\malian}[1]{{\color{blue}[\textbf{\sc malian}: \textit{#1}]}}
\newcommand{\dmin}[1]{{\color{green}[\textbf{\sc dmin}: \textit{#1}]}}
\else
\newcommand{\nskim}[1]{}
\newcommand{\malian}[1]{}
\newcommand{\dmin}[1]{}
\fi

\usepackage{xspace}
\usepackage{subfig}

\def\loadgen{\texttt{EtherLoadGen}\xspace}

\usepackage[linesnumbered,ruled]{algorithm2e}

\def\fwd{\texttt{L2Fwd}\xspace}

\usepackage{listings}
\usepackage{xcolor}

\definecolor{codegreen}{rgb}{0,0.6,0}
\definecolor{codegray}{rgb}{0.5,0.5,0.5}
\definecolor{codepurple}{rgb}{0.58,0,0.82}
\definecolor{backcolour}{rgb}{0.95,0.95,0.92}

\lstdefinestyle{mystyle}{
    backgroundcolor=\color{backcolour},   
    commentstyle=\color{codegreen},
    keywordstyle=\color{magenta},
    numberstyle=\tiny\color{codegray},
    stringstyle=\color{codepurple},
    basicstyle=\ttfamily\footnotesize,
    breakatwhitespace=false,         
    breaklines=true,                 
    captionpos=b,                    
    keepspaces=true,                 
    numbers=left,                    
    numbersep=5pt,                  
    showspaces=false,                
    showstringspaces=false,
    showtabs=false,                  
    tabsize=2
}

\RequirePackage[normalem]{ulem} 
\RequirePackage{color}\definecolor{RED}{rgb}{1,0,0}\definecolor{BLUE}{rgb}{0,0,1} 

\begin{document}
%


\title{Enabling Kernel Bypass Networking on gem5}
%
%
%
%

\author{\small{Siddharth Agarwal, Minwoo Lee, Ren Wang, and Mohammad Alian}

\IEEEcompsocitemizethanks{
\IEEEcompsocthanksitem 

S. Agarwal is with the Department of Electrical and Computer Engineering, University of Illinois, Urbana-Champaign, IL, USA. E-mail: sa10@illinois.edu.


\IEEEcompsocthanksitem 
R. Wang is with Intel Labs, OR, USA. E-mail: ren.wang@intel.com.

\IEEEcompsocthanksitem 
M. Lee and M. Alian are with the Department of Electrical Engineering and Computer Science, University of Kansas, KS, USA. E-mail: alian@ku.edu.

}
}

\IEEEtitleabstractindextext{%

\begin{abstract}
    Full-system simulation of computer systems is critical to capture the complex interplay between various hardware and software components in future systems. Modeling the network subsystem is indispensable to the fidelity of the full-system simulation due to the increasing importance of scale-out systems. The network software stack has undergone major changes over the last decade, and kernel-bypass networking stacks and data-plane networks are rapidly replacing the conventional kernel network stack. Nevertheless, the current state-of-the-art architectural simulators still use kernel networking which precludes realistic network application scenarios. In this work, we enable kernel bypass networking stack on gem5, the state-of-the-art full-system architectural simulator. We extend gem5's NIC hardware model and device driver to enable the support for userspace device drivers to run the DPDK framework. We also implement a network load generator hardware model in gem5 to generate various traffic patterns and perform per-packet timestamp and latency measurements without introducing packet loss. Our experimental results show that DPDK's simulated network bandwidth scales with the number of cores and NIC ports. As two use cases, we analyze the sensitivity of (1) network performance to several microarchitectural parameters, and (2) direct cache access (DCA) technology to DPDK burst size.  
    
    \end{abstract}

\begin{IEEEkeywords}
Network, Kernel Bypass, DPDK, gem5
\end{IEEEkeywords}}

\maketitle

\IEEEdisplaynontitleabstractindextext

\IEEEpeerreviewmaketitle

\section{Introduction}
\label{sec:introduction}
\IEEEPARstart{T}he evolution of networking technology enabled hundreds of gigabytes per second inter-server data transmission rates in datacenters, and terabit per seconds network interfaces are on the horizon~\cite{wade2020teraphy}. Proper handling of such high network rates in the processor microarchitecture and memory hierarchy is necessary to deliver high-quality end-to-end performance for emerging exascale applications. Unfortunately, the current tools for modeling and evaluating future system architectures have outdated models for the networking subsystem and at most can deliver several tens of gigabits per second network data rates to the processor and memory hierarchy. For instance, gem5's baseline network interface card (NIC) model only delivers $\sim$10Gbps network bandwidth with a single NIC running \texttt{iperf} TCP test ($\sim$20Gbps with 4 NICs). FireSim, which is an FPGA-based cycle-accurate simulator, supports networked simulation that delivers $\sim$1Gbps bandwidth per NIC~\cite{karandikar-2018-firesim}. 

With such low network bandwidth, the architectural simulation cannot be used for evaluating future terabit per second networked systems. In this work, we filled this gap in the gem5 simulator by enabling the DPDK software stack on gem5 that bypasses the Linux network software stack and delivers the maximum network bandwidth that a given processor architecture and memory hierarchy can sustain. In other words, we enable full-system gem5 to simulate networked systems in which the network stack is no longer the bottleneck; instead, in our setup, the processor and memory are the bottlenecks in network packet delivery. 

\begin{figure}[t]
\begin{center}
\includegraphics[width=\columnwidth]{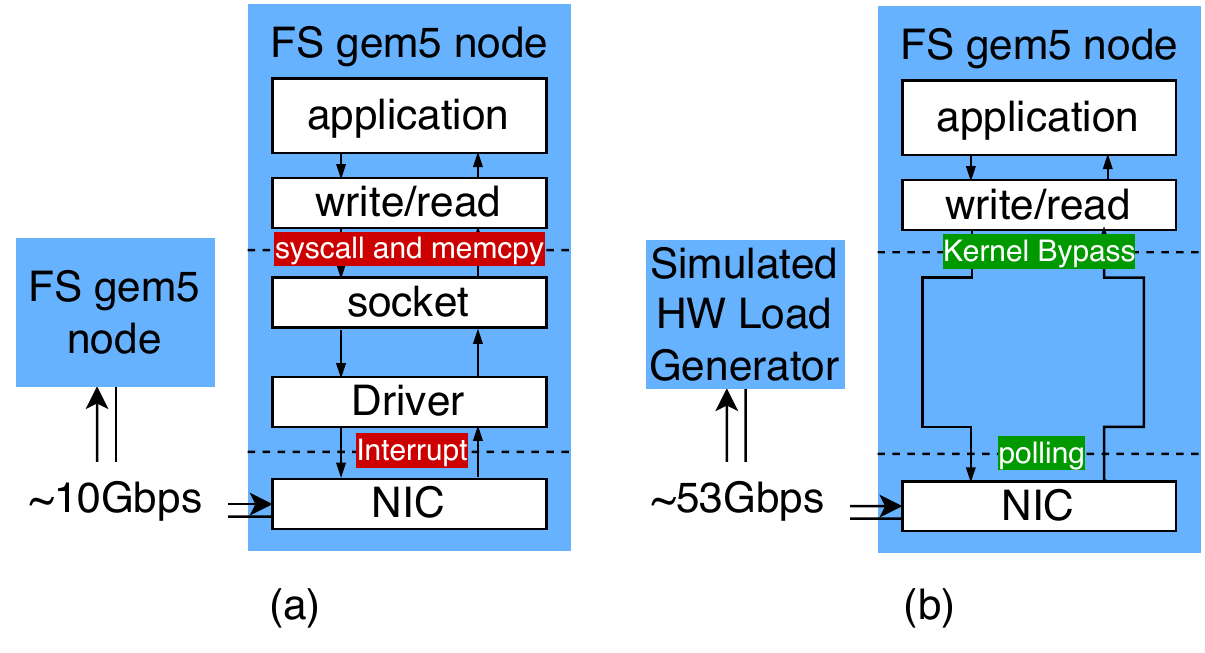}
\vspace{-19pt}
\caption{ (a) Baseline dual-mode gem5 with Linux kernel software stack for evaluating networked systems, (b) what is enabled in this work: kernel bypass software stack with hardware load generator model in gem5.}
\vspace{-20pt}
\label{fig:overview}
\end{center}
\end{figure}

In this paper, we describe the shortcomings of gem5's current network stack in evaluating future networked systems and explain our extensions to the gem5 hardware model and Linux kernel to enable a DPDK software stack with a polling mode driver (PMD) network interface. We also add a network traffic generator hardware model to gem5 that can be used to inject packets to the simulated network with configurable rate, packet size, and traffic pattern. The hardware traffic generator is widely used in the industry to stress test the network subsystem without experiencing any packet loss. Our experimental results show that the bandwidth of our DPDK PMD interface scales with the number of simulated network ports and processor cores. Each simulated network port pinned to a simulated ARM core -- loosely modeled after an ARM Cortex-A72 core -- in a quad-core setup can sustain network receive and transmit bandwidth of $\sim$25Gbps when running L2fwd DPDK application without any packet loss. As illustrated in Fig.\ref{fig:overview}, a single NIC port running \fwd DPDK application sustains $\sim$53Gbps while iperf only sustains $\sim$10Gbps.  
The complete simulation setup and extensions are open sourced in the following git repository 
\url{https://github.com/agsiddharth/CAL-DPDK-GEM5}. The authors are also working on integrating the changes to the vanilla release of gem5. 

\vspace{-10pt}
\section{Background and Motivation}
\label{sec:motivation}

Network packet processing in the Linux kernel suffers from the following bottlenecks: frequent system calls for packet transmission and reception, frequent buffer copies within kernel software stack and between kernel and userspace buffers, and long-latency interrupt processing and notification. Kernel bypass software stacks alleviate these bottlenecks by (1) reducing context switches between userspace and kernel space; (2) using large buffer allocations, huge pages, and zero-copy interfaces to reduce buffer management and data movement overheads; (3) using polling for RX and TX completion notification. 
Many kernel bypass protocols have been proposed and implemented, including Data Plane Development Kit (DPDK)~\cite{intelDPDK}. 
%

%
DPDK provides a userspace API for application developers. It reserves pinned hugepages and allows the NIC to DMA directly into the application's buffers. Since it is polling based, it eliminates context switching overheads as well.
%
DPDK application can be implemented in two modes: 

\noindent\textit{Run to completion} mode where the packet processing loop is: 
(1) retrieve RX packets through polling mode driver (PMD) RX API, 
(2) process packets on the same logical core,
(3) send pending packets through PMD TX API

\noindent \textit{Pipeline mode}, which lets cores pass packets between each other via a ring buffer to process packets.

\noindent \textbf{Current gem5 Network Stack.}


The current NIC simulation object in gem5 
loosely models Intel\textsuperscript{\textregistered} 8254x NIC series at a minimal functional level. 
Fig.~\ref{fig:overview}(a) shows a two-node full-system simulated system connected through the simulated NIC and Ethernet Link. Default gem5 uses Kernel interrupt-driven network stack and can only sustain up to $\sim$20Gbps network bandwidth using four powerful, multi-core, O3 ARM cores running at 3GHz frequency (See Sec.~\ref{sec:eval} for detailed experimental methodology). Such low network bandwidth does not sufficiently exercise the hardware and software stack of future networked systems modeled with gem5. Thus, the current gem5 model is not useful for evaluating networked systems supporting hundreds of gigabits per second network throughput. 


\textbf{Hardware Traffic Generators.}
One of the main concerns when evaluating networked systems is to load the system-under-test with real traffic and measure network bandwidth and per-packet, round-trip network latency without introducing extra latency or packet drop a the load generator node. 
The practice in the industry is to use hardware traffic generators that utilize FPGA line cards to generate packets with configurable traffic patterns, sizes, and protocols and provide detailed network statistics per transmitted and received packets~\cite{keysightloadgen}.



\vspace{-12pt}
\section{Linux Kernel Bypass in gem5}
\label{sec:arch}

This section discusses the changes we made to gem5 and DPDK framework to enable Kernel bypass networking and implement the hardware load generator model in gem5. We do not make any changes in the Linux kernel. 

\begin{figure}[t]
\begin{center}
\includegraphics[width=0.8\columnwidth]{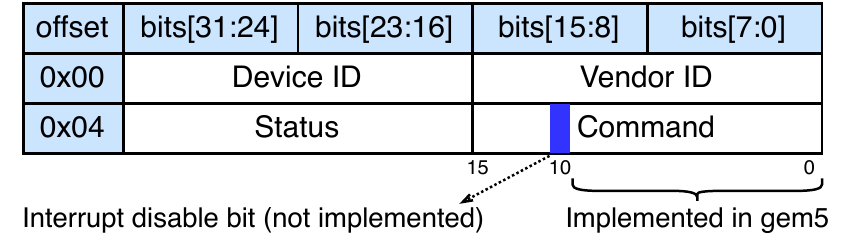}
\caption{ First 8 bytes of the PCI configuration space that includes PCI Command Register.}
\vspace{-10pt}
\label{fig:pcieconfig}
\end{center}
\end{figure}

\subsection{Changes to gem5}
\label{sec:arch:gem5changes}

The changes in gem5 are limited to PCI model to enable Userspace I/O (UIO) driver and the NIC model to enable byte granular PCI configuration space accesses. 

\subsubsection{Enable Userspace I/O Driver}
\texttt{uio\_pci\_generic} driver in Linux enables a userspace application to directly access the address space of a PCI device. 
DPDK uses this driver to gain the userspace application access to the PCI config space and implement a polling mode driver. 
The default gem5 does not enable the \texttt{uio\_pci\_generic} driver during boot as the PCI Command Register is not fully implemented in gem5. Fig.\ref{fig:pcieconfig} shows the first 8 bytes of the PCI configuration space that includes the 16-bit Command Register at offset \texttt{0x04}. The baseline gem5 implements bits 0-9 of the Command Register but does not implement bit-10, which is the \textit{interrupt disable} bit. 
We implement \textit{interrupt disable} bit in gem5 PCI model so the Linux kernel can disable the interrupts for the PCI devices on gem5 which is necessary to support \texttt{uio\_pci\_generic} driver. 

\subsubsection{Enable Byte-Granular Access to PCI Configuration Space}

The default gem5 only supports 16-bit accesses to the Command Register shown in Fig.\ref{fig:pcieconfig}. In fact, this is rational since the size of the Command register is 16-bits. However, we observed that DPDK accesses the Command register using 8-bit memory accesses. Such byte-granular accesses are being ignored in gem5, and therefore DPDK cannot properly read and write the upper half of the Command Register (offset \texttt{0x05} of the PCI config space). We extended the \texttt{readConfig} and \texttt{writeConfig} functions in the gem5 PCI model to enable byte-granular accesses to the Command Register.

\subsubsection{Implement Interrupt Mask Register in the NIC model}

The last modification in gem5 is to implement Interrupt Mask Register in the \texttt{i8254xGBe} device model. Interestingly, this register is included in the \texttt{i8254xGBe} model, but the read and write methods for accessing the register are not implemented in the current gem5 release. We implemented the read and write methods to enable DPDK to launch its polling mode driver.

\subsubsection{Enable the NIC Model to Correctly Operate with a PMD}

%

NIC devices keep a handful of available descriptors (usually 32$\sim$64 descriptors) that can be populated upon receiving a packet on an on-chip cache which is called \textit{descriptor cache}. Descriptor cache improves the performance as the NIC does not need to fetch available descriptors from the CPU memory on demand. 
The NIC gradually writes back the descriptor cache to the CPU memory (using DMA), and then the CPU is notified of received packets.   

The current gem5's NIC model writes back the received descriptors based on a threshold set by the Linux kernel. Once the number of used descriptors is higher than a threshold, NIC initiates a writeback. %
When using a PMD, the threshold registers in the NIC model are not properly set, and thus the NIC starts writing back the descriptors when all of them are used. This means that packets are DMAed to the CPU memory in large batches (32$\sim$64 packets), which causes unrealistic pressure on the CPU memory subsystem and increases the possibility of packet drops at high receive rates. 
We implemented a parameter for the NIC where the user can control the threshold of descriptor writebacks in gem5.


\subsection{Changes to DPDK}
\label{sec:arch:dpdkchanges}

The DPDK Environment Abstraction Layer (EAL) relies on vendor ID checks to match a device and a PMD. We modify the DPDK source to skip these checks and force the matching of the gem5 device to the e1000 PMD.
Unmodified DPDK cannot fetch the correct vendor ID when running on gem5 and therefore fails to call the proper PMD driver. 
We suspect this is because some manufacturer-specific information is missing in the gem5 NIC model. 
Note that skipping the vendor ID test does not have any impact on gem5 simulations as the current gem5 release has only the e1000 NIC model. If new NIC models are added to gem5, the DPDK framework should be recompiled after hard-coding the PMD driver to use a different NIC model.

\subsection{Hardware Load Generator Model.}
\label{sec:arch:loadgen}

The hardware load generator model can generate packets at arbitrary rates and sizes. We implement a simulation object called \loadgen that has a single Ethernet port and can directly connect to the NIC port of a simulated node as shown in Fig.\ref{fig:overview}(b). Therefore, for simple network benchmarking, one does not need to run distributed or dual mode gem5 simulations and a single system simulation is enough. This significantly improves the simulation speed. 
The parameters of \loadgen are packet rate, packet size, and protocol. They can be statically set while launching a simulation or a packet trace can be passed to the simulator to be replayed by \loadgen. In the static mode, \loadgen create Ethernet packets with the specified size and send them at a fixed rate to the Ethernet port. The static protocol that we support for now is plain Ethernet packets. If a trace file is provided, \loadgen will read from the trace file and generate traffic based on the timestamps, sizes, and protocol in the trace file. 

\loadgen adds a timestamp to each outgoing packet at a configurable offset and compares the timestamp with the current tick on incoming packets to compute per-packet round-trip latency. 
\loadgen reports mean, median, standard deviation, and tail latency of network packets in the statistics file. 
It also produces a packet drop percentage and a histogram of packet forwarding latency. The load generator model enables simple network benchmarking in gem5 without the need to simulate multiple system nodes. This is similar to the practice in the industry for using hardware load generators to evaluate the performance of the network~\cite{spirentAION}.

\loadgen also supports a bandwidth test mode where it gradually increases the bandwidth to find the maximum sustainable bandwidth of a server, which is the maximum bandwidth that a server can sustain without packet drops.

\section{Evaluation}
\label{sec:eval}

\begin{table}[t]
\caption{Simulation configuration. 
}
\centering
\resizebox{0.48\textwidth}{!}{
{
\begin{tabular}{lcc}
\toprule
Parameters & Baseline Values \\
\midrule
    Core freq: & 2GHz             \\
    Superscalar     &  3 ways      \\
    ROB/IQ/LQ/SQ entries   & 384/128/128/128 \\
    Int \& FP physical registers & 128 \& 192 \\
    Branch predictor/BTB entries       & BiMode/2048 \\
    Caches (size, assoc): I/D/L2  & 32KB,2/64KB,2/2MB,16ways \\
    L1I/L1D/L2 latency,MSHRs   & 1/2/12 cycles, 2/6/16 MSHRs \\
    DRAM/mem size               & DDR4-3200-8x8/2GB \\
    iocache   & 24 cycles, 16 MSHRs \\
    Network latency/BW   & 1$\mu$/200Gbps \\
    DPDK   & Version 20.11.3\\
    Operating system     & Linux Linaro (kernel 5.4.0) \\
    gem5 version     & v21.1.0.2 \\
\bottomrule
\end{tabular}
}
}
\label{table:proc_config}
\end{table}

\subsection{Methodology}
\label{sec:eval:method}
Table~\ref{table:proc_config} shows the baseline gem5 configuration we used for the experiments. We use \texttt{iperf} and \fwd for Kernel and DPDK experiments. In this section, we first explain how to build DPDK on gem5 full-system disk-image and write scripts to run a simple \fwd DPDK application and load it with the simulated hardware packet generator.
Then we show the results that validate the correctness of our Kernel bypass setup. Next, we use \texttt{iperf} and \fwd to compare the scalability of Kernel stack and DPDK on gem5. Lastly, we perform a sensitivity analysis on microarchitecture configurations for network bandwidth when using Kernel stack and DPDK.


To build a kernel and disk image for gem5, we use the buildroot tool. The Kernel needs to be compiled with support for huge pages, and the kernel module \texttt{uio\_pci\_generic}. Listing~\ref{listing:kernconfig} shows the Kernel config option needed to be enabled in buildroot tool for DPDK.

\begin{lstlisting}[language={Bash}, caption={Kern CONFIG options needed for DPDK.},label={listing:kernconfig}]
CONFIG_HUGETLBFS=y
CONFIG_HUGETLB_PAGE=y
CONFIG_UIO=y
CONFIG_PCI=y
CONFIG_UIO_PCI_GENERIC=m

\end{lstlisting}

\begin{lstlisting}[language={Bash}, caption={Bash script for setting up the environment for running DPDK applications on gem5.},label={listing:bash}]
 modprobe uio_pci_generic
 dpdk-devbind.py -b uio_pci_generic 00:02.0
 echo 2048 > /sys/kernel/mm/hugepages/hugepages-2048kB/nr_hugepages
 dpdk-testpmd


\end{lstlisting}

Listing~\ref{listing:bash} shows the bash script for bringing up the Userspace IO (UIO) driver (line 1), binding it to a NIC port (line 2), allocating huge pages, and lastly, starting the \textit{testpmd} application. As shown in the listing, the procedure for running DPDK applications on gem5 is identical to running DPDK apps on bare metal hardware. 


\vspace{-10pt}
\subsection{Experimental Results}
\label{sec:eval:expr}

In this section, we show experimental results that verify the functionality of DPDK enabled hardware/software stack on gem5 
and illustrate the scalability of our kernel bypass network software stack on gem5. Since validating gem5's performance is out of the scope of this paper, we do not compare the network bandwidth of a bare metal system with our gem5 setup and only report the performance of gem5. There are on-going efforts to validate gem5 performance for different ISAs~\cite{gem5-validation}. 

To verify DPDK's functionality on the simulated system, we modify \fwd sample application to print the content of the packets received from the network. We always receive the correct content regardless of the packet size and network configuration. This experimentally verifies the correct functionality of the kernel bypass stack on gem5. 
%

%



\begin{figure}[t]
\begin{center}
\includegraphics[width=0.9\columnwidth]{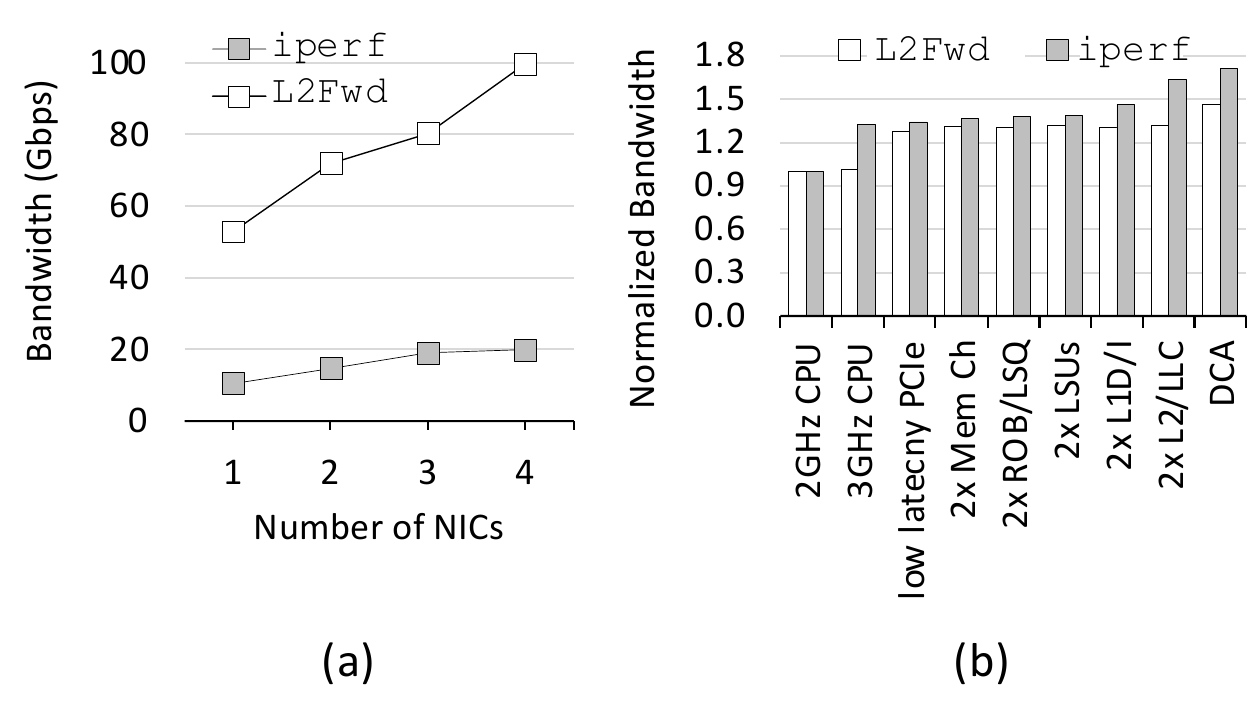}
\vspace{-5pt}
\caption{Linux kernel and DPDK userspace software stack (a) scalability and (b) sensitivity to architectural parameters.}
\vspace{-20pt}
\label{fig:scalability}
\end{center}
\end{figure}

Fig.~\ref{fig:scalability}(a) shows the \textit{maximum sustainable bandwidth} of a single gem5 node when equipped with up to 4 NICs, running \texttt{iperf} and \fwd. 
We define the \textit{maximum sustainable bandwidth} as the maximum bandwidth without packet drop. As explained in Sec.\ref{sec:arch:loadgen}, \loadgen supports a mode where it gradually increases the packet rate and monitors the responses received from the server to find the maximum sustainable bandwidth.
Fig.\ref{fig:scalability}(a) compares the bandwidth achieved using Linux kernel stack (\texttt{iperf} configuration) and DPDK (\fwd configuration).
Fig.\ref{fig:scalability}(a) has two highlights: (1) DPDK delivers much higher bandwidth compared with kernel stack. As shown in the figure, \fwd sustains 5.4$\times$ and 4.9$\times$ more bandwidth compared with \texttt{iperf} using 1 and 4 NICs, respectively. (2) adding more NICs scales DPDK's network bandwidth better than the Linux kernel. As shown in the figure, moving from 3 NICs to 4 NICs, the DPDK software stack has 24.1\% higher bandwidth, while the Linux kernel stack only sustains 5.3\% higher network bandwidth. 

%

\vspace{-10pt}
\section{Example Use Cases}

\subsection{Micro Architectural Sensitivity Analysis}

In this subsection, we analyze the sensitivity of the network bandwidth to various microarchitectural configurations. We perform the analysis for both Linux kernel and kernel bypass network stacks and compare the results. We start with a baseline node configuration shown in Table~\ref{table:proc_config} (\texttt{2GHz CPU} configuration in Fig.\ref{fig:scalability}(b)). Then increase the CPU frequency (\texttt{3GHz CPU}), reduce the PICe bus latency (\texttt{low latency PCIe}), double the number of memory channels (\texttt{2x Mem Ch}), double the size of ROB and LSQ (\texttt{2xROB/LSQ}), double the number of load-store functional units in the processor pipeline (\texttt{2xLSUs}), double the size of L1 data and instruction caches (\texttt{2xL1D/I}), double the size of L2 and LLC caches (\texttt{2xL2/LLC}), and lastly enable direct cache access~\cite{alian-2020-ddio} to place received network data on the LLC instead of DRAM (\texttt{DCA}). Fig.\ref{fig:scalability}(b) shows the maximum sustainable bandwidth for all the mentioned configurations. 
Note that the enhancements are accumulative, meaning that we apply each optimization on top of the previous one, i.e., the \texttt{DCA} configuration runs at 3GHz, has low latency PCIe, twice the number of memory channels, twice ROB, LSQ, L1D, L1D, L2, LLC sizes of the configuration listed in Table~\ref{table:proc_config}.

As shown in Fig.\ref{fig:scalability}(b), different microarchitectural parameters impact DPDK and Linux kernel stacks differently. For example, increasing CPU frequency alone improves DPDK bandwidth by 1.2\% but improves Linux kernel stack bandwidth by 32.5\%. This is due to the CPU intensity of the kernel stack compared to the userspace DPDK stack. %
Fig.\ref{fig:scalability}(b) clearly shows that conducting architectural research using an old software stack can lead to incorrect assumptions and optimizations.

\vspace{-10pt}
\subsection{Sensitivity of Direct Cache Access Performance to Burst Size}

\begin{figure}[t]
\begin{center}
\includegraphics[width=0.8\columnwidth]{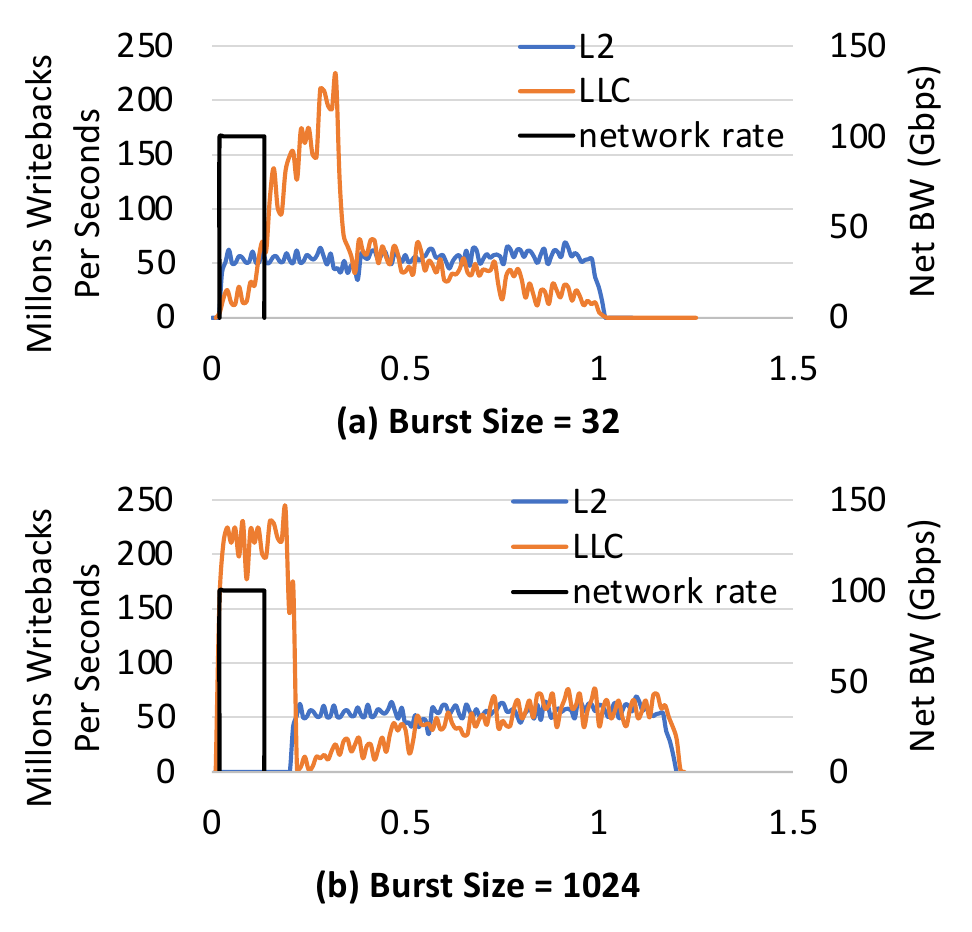}
\vspace{-5pt}
\caption{Sensitivity of L2 and LLC writebacks to the packet processing batch size.}
\vspace{-20pt}
\label{fig:sen-writeback}
\end{center}
\end{figure}

As another usecase for the userspace networking and network load generator, Fig.\ref{fig:sen-writeback} plots the impact of \fwd burst size on the writeback rate of L2 and L3 caches when receiving 1024 packets in a short time interval. Fig.\ref{fig:sen-writeback}(a) \fwd aggregate forwarding in burst of 32 packets while in Fig.\ref{fig:sen-writeback}(b), \fwd waits until 1024 packets are received and then start the forwarding. The simulated node implements a non-inclusive L2 with DCA enabled. 
As shown, a large batch size results in LLC contention at the beginning of the burst arrival as many packets will be DMA transferred to the LLC (in the ring buffer) within a short interval. A shorter batch size overlaps the processing of the received packets with the DMA from NIC to LLC and since L2 is non-inclusive, demand misses from L2 make space in the LLC for incoming packets; therefore, Fig.\ref{fig:sen-writeback}(a) has lower LLC writeback rate.

\vspace{-15pt}
\section{Conclusion}
\label{sec:conclusion}

In this paper, we introduced gem5's userspace networking stack. We explained the changes we made to enable gem5 to run DPDK applications. 
We show that the bandwidth of \fwd DPDK application running on gem5 follows the same trend when running on bare metal hardware. We showed that the bandwidth of DPDK applications running on gem5 scales significantly better than applications that use the Linux kernel stack. Using gem5 running \texttt{iperf} and \fwd, we performed a sensitivity analysis on microarchitecture optimizations and showed that kernel bypass network applications are sensitive to different microarchitecture optimizations compared with Linux kernel network applications. 
We released the source code, scripts, and instructions to create disk images, install DPDK, and run DPDK applications in the following git repository \url{https://github.com/agsiddharth/CAL-DPDK-GEM5}..

\vspace{-10pt}

\section*{Acknowledgement}
\vspace{-0.5em}
   This research was in part supported by an NSF grant (CNS-2213807).
\vspace{-1.5em}

\ifCLASSOPTIONcaptionsoff
  \newpage
\fi

\bibliographystyle{plain}
\bibliography{ref-abbrev}

\end{document}